\documentclass[aps, prd, superscriptaddress, showkeys, showpacs, twocolumn, longbibliography, nofootinbib]{revtex4-1}  
\usepackage[utf8]{inputenc}
\usepackage{amsmath}
\usepackage{amsfonts}
\usepackage{amsthm}
\usepackage{amssymb}
\usepackage{float}
\usepackage{braket}
\usepackage{graphicx}
\usepackage{url}
\usepackage[colorlinks=true, pdfstartview=FitV, linkcolor=red, citecolor=blue, urlcolor=blue]{hyperref}
\usepackage{slashed}
\usepackage[normalem]{ulem}
\newcommand{\frmn}{\hat{a}}
\newcommand{\anfm}{\hat{c}}

\begin{document}
\title{ Applicability of kinetic theory in strongly coupled thermal quantum systems}

\author{Shile Chen}
\email[]{shchen@lns.infn.it}
\affiliation{Laboratori Nazionali del Sud, INFN-LNS, Via S. Sofia 62, I-95123 Catania, Italy}
\affiliation{Department of Physics, Tsinghua University, Beijing 100084, China}

\author{Shuzhe Shi}
\email[]{shuzhe-shi@tsinghua.edu.cn}
\affiliation{Department of Physics, Tsinghua University, Beijing 100084, China}
\affiliation{State Key Laboratory of Low-Dimensional Quantum Physics, Tsinghua University, Beijing 100084, China}

\author{Pengfei Zhuang}
\email[]{zhuangpf@tsinghua.edu.cn}
\affiliation{Department of Physics, Yantai University, Yantai 264005, China}
\affiliation{Department of Physics, Tsinghua University, Beijing 100084, China}
\affiliation{South Center for Nuclear-Science Theory, Institute of Modern Physics, Chinese Academy of Sciences, Huizhou 516000, China}

\begin{abstract} 
In this work, we construct one-dimensional interacting lattice spinor theories with discretization in momentum space. We focus on strongly interacting Schwinger and Nambu--Jona-Lasinio models and perform ab-initio calculation of their single-particle and two-particle momentum distribution functions at finite temperature. We observe, at low temperature, high-momentum tail in single-particle and two particle distribution which reveals relative momentum in fermion-antifermion boundstates, as well as quasi-free spinor gases behavior at high temperature. The non-vanishing connected four-momentum function reveals the quantum coherence in momentum space under thermal equilibrium of the system and indicate the single particle correlation would remember more microscopic details within a thermal system. Overall, for a high-enough temperature at which the thermal kinetic energy comparable with the interaction, we observe that the two-particle correlation is subdominant compared to the single particle distributions, which indicates the applicability of kinetic theory. 
\end{abstract}
\maketitle

\section{Introduction}

Understanding thermalization of strongly coupled quantum matter far from equilibrium is a profound challenge in modern physics. The hot and dense medium created in relativistic heavy-ion collisions, a system governed by strong interaction, is a unique example. These collisions create quark-gluon plasma(QGP), a state of matter that the color degrees of freedom become deconfined. To describe the evolution of such a complex system in the framework of quantum chromodynamics(QCD) with high degrees of freedom, many effective theories are adopted, among them the hydrodynamics~\cite{Shuryak:2014zxa, Shuryak:2003xe} and semi-classical kinetic equation~\cite{Baier:2000sb, Kurkela:2015qoa, Blaizot:2011xf, Blaizot:2013lga, Berges:2012us, Berges:2014bba, Blaizot:2014jna, Xu:2014ega, Lu:2025yry, Lu:2025asx} play important roles in such a circumstance. However, the great success of hydrodynamic and kinetic theory in describing the QGP evolution from a very early stage far from thermal equilibrium, raise another question: does this strongly interacting system already reduce to the low dimensional manifold description within the hydrodynamic observables for the former one and does the single particle representation efficient enough to include all the dominant dynamics of the system for the latter. The latter is particularly under question, because semiclassical kinetic theories usually treat interaction perturbatively so that multi-particle correlations are assumed to be negligible. Demonstrating the applicability of such classical/semi-classical effective theories requires verification in an ab-initio manner. 

In the recent studies, researches have adopted the quantum simulation method within a QCD-like, one-dimensional quantum electrodynamic theory -- Schwinger model~\cite{Schwinger:1962tp} to study the thermalization and hydrodynamization process in a strongly interacting system and trying to understand this fast reduce of degrees of freedom within a unitary time evolution of a quantum state. Despite its simplicity, it captures essential features of QCD, including confinement, chiral symmetry breaking, and a non-trivial phase structure depending on mass-to-coupling ratio and a topological $\theta$-angle~\cite{Lowenstein:1971fc, Jayewardena:1988td, Smilga:1992hx, Adam:1993fc, Adam:1997wt, Coleman:1976uz, Adam:1995us, Adam:1996np, Adam:1996qk}. Under this many body framework, the thermalization and hydrodynamization signals could be extracted from the quantum state evolution~\cite{Chen:2024pee, Shao:2025obi, Shao:2025ygy}. In the semi-classical kinetic theories that always assume perturbative interactions, the multiparticle joined distribution can be treated as the production of few-particle distributions~\cite{2024arXiv240807818D, 2025arXiv250301800D}, but this has not been verified for strongly-coupled quantum systems. In recent studies, many phenomena like scattering amplitude coherence and energy-energy correlation in jet physics reveal that the quantum information in momentum space is stored beyond single-particle distribution~\cite{Song:2025bdj, Gao:2025evv, Barata:2025zku}. It would be more straightforward to study the momentum-related quantities with numerical simulation of quantum fields in momentum space.

In this work, we focus on the one dimensional models including Schwinger model and Nambu--Jona-Lasinio (NJL) model. We discretize them in momentum space to investigate the thermal one particle momentum distribution and momentum correlation in a finite temperature quantum system. Noting the field theories are typically non-local in momentum space, the applicability of tensor network method is not secured. Yet, the total momentum in such representation is strictly conserved, which makes the Hamiltonian block-diagonalized by construction and allows us adopt the exact diagonalization for intermediate-size systems. 

The rest of the paper is organized as follows. In Sections~\ref{sec:momentum_rep} we start from the second quantization to build and discretize two field theories in momentum space. Section~\ref{sec:results1} and ~\ref{sec:results2} present the simulation results of single particle distribution in comparison with interaction-free case and the momentum-momentum correlation respectively. Section~\ref{sec:summary} is the summary and outlook of our work.

\section{Spinor theories on momentum lattices}\label{sec:momentum_rep}
Current quantum computations---including simulations on both quantum and classical devices---of lattice field theories almost exclusively focus on discretization in coordinate space. Na\"ive discretization of the spinor field suffers from the fermion doubling problem~\cite{Nielsen:1981hk}. The staggered fermion scheme, which distributes spinor components over different lattice sites, removes the doublers, but it breaks the exact symmetry under parity and charge conjugation transformations. Another solution is the Wilson fermion, which adds a momentum‑dependent mass term (the Wilson term) to the action. This term gives the doubler modes a mass proportional to the inverse lattice spacing, decoupling them in the continuum limit. However, the Wilson term explicitly breaks chiral symmetry, resulting in an additive mass renormalization and $\mathcal{O}(a)$ lattice artifacts that slow the approach to the continuum. Additionally, in schemes with coordinate‑space discretization, the definition of momentum‑related operators---usually defined via a Fourier transformation over the finite volume---suffers from a number of subtleties. Momenta are restricted to a discrete set determined by the box size, which limits the accessible kinematic points and introduces systematic uncertainties from boundary effects. The hypercubic lattice breaks continuous rotational symmetry, causing operator mixing and complicating renormalization. Moreover, the oscillating phase factors in the Fourier sum can amplify statistical fluctuations, leading to a poor signal‑to‑noise ratio for high‑momentum observables. These challenges can be circumvented by formulating the field theory directly on a momentum lattice, where fermion doubling is avoided without spoiling chiral symmetry and momentum‑space operators are local. In this work, we construct lattice spinor field theories in momentum space and perform ab initio simulations to study the finite‑temperature properties of the momentum distribution in interacting spinor theories. To the best of our knowledge, simulations of lattice field theories in momentum space have been performed only in the Euclidean‑time lattice framework, most notably for the massive Schwinger model~\cite{Kroger:1993jd, Kroger:1996hj, Kroger:1998se}.

To construct the lattice fermion theories in the momentum space, we first rewrite the spinor fields to momentum space after quantization,
\begin{align}
\begin{split}
\psi(z) =\;&
    \int \frac{dp}{2\pi\sqrt{2E_p}}\Big(
    u(p) \frmn^{}(p) e^{i p z} +  v(p) \anfm^\dagger(p) e^{-i p z}
    \Big)\,,\\
\bar{\psi}(z) =\;&
    \int \frac{dp}{2\pi\sqrt{2E_p}}\Big(
    \bar{u}(p) \frmn^\dagger(p) e^{-i p z} +  \bar{v}(p) \anfm^{}(p) e^{i p z}
    \Big)\,,
\end{split}
\end{align}
with $E_p = \sqrt{p^2+m^2}$, $\anfm^\dagger(p)$ and $\anfm^{}(p)$ are, respectively, the creation and annihilation operators for an antifermion with momentum $p$ and likewise $\frmn^\dagger(p)$ and $\frmn^{}(p)$ are for fermions. The only non-vanishing anticommutation relations are
\begin{align}
    \{\frmn^{}(p), \frmn^{\dagger}(k)\} 
    = \{\anfm^{}(p), \anfm^{\dagger}(k)\}
    = 2\pi\,\delta(p-k).
    \label{eq:anticommutation}
\end{align}
Under the convention $\gamma^0 = \Big(\begin{matrix}1 & 0 \\0 & -1\end{matrix}\Big)$, $\gamma^1 = \Big(\begin{matrix}0 & 1 \\-1 & 0\end{matrix}\Big)$ and $\gamma^5 = \Big(\begin{matrix}0 & 1 \\1 & 0\end{matrix}\Big)$, the wavefunctions read
\begin{align}
u(p) = 
\left(\begin{array}{c}
    \sqrt{E_p+m} \\
    p/\sqrt{E_p+m}
\end{array}\right)\,,\quad
v(p) = 
\left(\begin{array}{c}
    p/\sqrt{E_p+m} \\
    \sqrt{E_p+m}
\end{array}\right)\,.
\end{align}

With this representation, it is clear that the free Dirac Hamiltonian is diagonalized in the momentum space,
\begin{align}
\begin{split}
\hat{H}_\mathrm{Dirac} \equiv\,&
    \int \big( \frac{i}{2} \big(\bar\psi' \gamma^1\psi  -\bar\psi \gamma^1 \psi' \big) + m \bar\psi \psi \big) \mathrm{d}z
\\=\;&
    \int \frac{E_p\,\mathrm{d}p}{2\pi}
    \big(\frmn^\dagger(p) \frmn^{}(p) + \anfm^\dagger(p) \anfm^{}(p) \big)\,,
\end{split}
\end{align}
and the total momentum and charge operators read, respectively,
\begin{align}
\begin{split}
\hat{P} \equiv\;& 
    \int (\psi^\dagger \psi'-\psi'^\dagger \psi) \frac{\mathrm{d}z}{2i}
=
    \int \frac{p\,\mathrm{d}p}{2\pi}\big(\frmn^\dagger(p) \frmn^{}(p) 
    + \anfm^\dagger(p)  \anfm^{}(p)
     \big)\,,\\
\hat{Q} \equiv\;& 
    \int \psi^\dagger \psi \,\mathrm{d}z
=
    \int \frac{\mathrm{d}p}{2\pi}\big(\frmn^\dagger(p) \frmn^{}(p)
    - \anfm^\dagger(p)  \anfm^{}(p)
     \big)\,.
\end{split}
\end{align}
We have denoted that $\psi'(z) \equiv \partial_z \psi(z)$ and similarly for $\psi'^\dagger$ and $\bar\psi'$.

In this work, we consider two interacting spinor theories, the NJL model and the massive Schwinger model, both of which are widely studied as toy models that mimics QCD.
The NJL model~\cite{Nambu:1961tp} introduces four-fermion interaction terms in the Hamiltonian, $\hat{H}_\mathrm{NJL} = \hat{H}_\mathrm{Dirac} + \hat{H}_\mathrm{NJL}^\mathrm{int}$. With Gross--Neveu scalar interaction, 
\begin{align}
\begin{split}
    \hat{H}_\mathrm{NJL}^\mathrm{int} = g_\mathrm{NJL}^2\int (\bar\psi \psi)^2 \mathrm{d}z,
\end{split}
\end{align}
the NJL model exhibits chiral phase transition at finite temperature, which is an important feature of QCD~\cite{Hatsuda:1994pi,Buballa:2003qv,Scavenius:2000qd,He:2005nk}. Such a transition is also observed in 1+1D lattice simulations~\cite{Czajka:2021yll, Zhang:2024fgv}. 

 In the Schwinger model the interaction between spinor fields is carried by a gauge field. In one dimension, the gauge field operators can be fully fixed by the spinor operators, upon a boundary term, because the Gauss' law, which ensures gauge invariance, determines electric field $\hat{\mathcal{E}}(z) = \mathcal{E}_\mathrm{bnd} + g_\mathrm{Sch} \int_{z_\mathrm{bnd}}^{z} \psi^\dagger(z') \psi(z') \mathrm{d}z'$. See e.g.~\cite{Chen:2024pee} for more details. 
Under this representation, the Schwinger Hamiltonian $\hat{H}_\mathrm{Sch} = \hat{H}_\mathrm{Dirac} + \hat{H}_\mathrm{Sch}^\mathrm{int}$ exhibits an interaction term coming from the electric field energy,
\begin{align}
\begin{split}
    \hat{H}_\mathrm{Sch}^\mathrm{int} = \frac{1}{2}\int \hat{\mathcal{E}}^2(z) \mathrm{d}z.
\end{split}
\end{align}

In momentum space, the four-fermion-interaction in the NJL model and the electric field energy in the Schwinger model respectively read 
\begin{align}
\begin{split}
    \hat{H}_\mathrm{NJL}^\mathrm{int} 
=
    g_\mathrm{NJL}^2\int \frac{\mathrm{d}q}{2\pi} \hat{\sigma}(q) \hat{\sigma}(-q),
    \label{eq:Hint_NJL}
\end{split}\\
\begin{split}
    \hat{H}_\mathrm{Sch}^\mathrm{int}
=
    g_\mathrm{Sch}^2\int \frac{\mathrm{d}q}{4\pi} \hat{e}(q) \hat{e}(-q),
    \label{eq:Hint_Sch}
\end{split}
\end{align}
where we have set the boundary condition of gauge field as $\mathcal{E}_\mathrm{bnd} = 0$ at ${z}_\mathrm{bnd} = -\infty$ and introduced the Fourier transformations of condensate [$\bar\psi(z)\psi(z)$] and electric field [$\hat{\mathcal{E}}(z)$], respectively, 
\begin{align}
\begin{split}
\hat{\sigma}(q) \equiv\,&
    \int\frac{\mathrm{d}p}{4\pi}\Big(S(\frac{q}{2}+p,\frac{q}{2}-p)
    \big(\frmn^\dagger(p+\frac{q}{2}) \frmn^{}(p-\frac{q}{2}) 
\\&\qquad\quad
    - \anfm^{}(p-\frac{q}{2})  \anfm^\dagger(p+\frac{q}{2}) \big)
\\&
    -A(\frac{q}{2}+p,\frac{q}{2}-p)
    \big(\frmn^\dagger(p+\frac{q}{2}) \anfm^\dagger(-p+\frac{q}{2}) 
\\&\qquad\quad
    - \frmn^{}(p-\frac{q}{2}) \anfm^{}(-p-\frac{q}{2})\big) \Big),
\end{split}\label{eq:q}\\
\begin{split}
\hat{e}(q) \equiv\,&
     \int\frac{\mathrm{d}p}{4\pi\,q}\Big( 
    S(p+\frac{q}{2}, p-\frac{q}{2}) 
    \big(\frmn^\dagger(p-\frac{q}{2}) \frmn^{}(p+\frac{q}{2}) 
\\&\qquad\quad
    - \anfm^\dagger(p-\frac{q}{2})  \anfm^{}(p+\frac{q}{2}) \big)
\\&
    -A(p+\frac{q}{2}, p-\frac{q}{2}) 
    \big(\frmn^\dagger(p-\frac{q}{2}) \anfm^\dagger(-p-\frac{q}{2}) 
\\&\qquad\quad
    + \frmn^{}(-p+\frac{q}{2}) \anfm^{}(p+\frac{q}{2})\big) \Big)\,,
\end{split}\label{eq:c}
\end{align}
with notations $S(p_a, p_b) \equiv \frac{p_a p_b + (m+E_a)(m+E_b)}{\sqrt{(m+E_a)(m+E_b)E_a E_b}}$ and $A(p_a, p_b) \equiv \frac{(m+E_b)p_a - (m+E_a)p_b}{\sqrt{(m+E_a)(m+E_b)E_a E_b}}$.

\begin{figure}[!hbpt]
    \centering
    \includegraphics[width=1.0\linewidth]{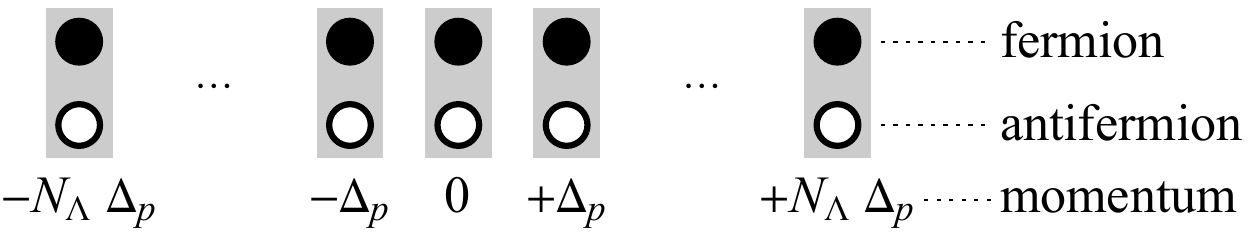}
    \caption{The sites layout in of momentum lattice. A filled (open) circle represents a fermion (antifermion) site. \label{fig:sites_layout}}
\end{figure}
In numerical simulations, we discretize the momentum space into grids with $p= \{ 0, \pm\Delta_p, \cdots \pm\,N_\Lambda\Delta_p\}$. We take the momentum Fock basis with $\ket{0}$ being the empty state and $\ket{1}$ representing the state that is occupied by either fermion or antifermion.
We map these 1D fermion theories on a one-dimensional spin chain illustrated by Fig.~\ref{fig:sites_layout}, where the odd and even sites represent fermions and antifermions, respectively.
For convenient comparison between discrete and continuous theories, we introduce a notation
\begin{align}
    \anfm^\dagger_{k} \equiv 
    \sqrt{\frac{\Delta_p}{2\pi}}\,
    \anfm^\dagger(p=k\Delta_p),
    \quad k\in \{0, \pm 1, \cdots \pm N_\Lambda\}.
\end{align}
Likewise for $\anfm^{}_{k}$, $\frmn^\dagger_{k}$, and $\frmn^{}_{k}$.
They can be represented by $\frmn_{j} = \chi_{2(N_\Lambda +j)+1}^{}$ and $\anfm_{j} = \chi_{2(N_\Lambda +j)+2}^{}$ so that the anti-commutation relations~\eqref{eq:anticommutation} are followed,
with $\chi_{n}^{}$ being the the Jordan--Wigner matrices~\cite{Jordan:1928wi} and they satisfy 
\begin{align}
\{ \chi_n^{}, \chi_{m}^{}\} = \{ \chi_n^{\dagger}, \chi_{m}^{\dagger}\} = 0,\quad
 \{ \chi_n^{}, \chi_{m}^{\dagger}\} = \delta_{nm}\,.
\end{align}

The free Dirac Hamiltonian, the total momentum, and the total charge operators read
\begin{align}
\begin{split}
\hat{H}_\mathrm{Dirac} 
=\;&
    \sum_{k=-N_\Lambda}^{N_\Lambda} \sqrt{m^2+k^2\Delta_p^2}\,
    \big(\frmn_{k}^\dagger \frmn_{k}^{} +  \anfm^\dagger_{k} \anfm_{k}^{} \big)\,,\\
\hat{P}
=\;&
    \sum_{k=-N_\Lambda}^{N_\Lambda} k\,\Delta_p\,
    \big(\frmn_{k}^\dagger \frmn_{k}^{} +  \anfm^\dagger_{k} \anfm_{k}^{} \big)\,,\\
\hat{Q}
=\;&
    \sum_{k=-N_\Lambda}^{N_\Lambda}
    \big(\frmn_{k}^\dagger \frmn_{k}^{} -  \anfm^\dagger_{k} \anfm_{k}^{} \big)\,.
\end{split}
\end{align}

Note that all spinor operators shall be defined on integer grids, $q$ in Eqs.~\eqref{eq:q} and~\eqref{eq:c} must be integers of $\Delta_p$, and $p/\Delta_p$ shall be either odd times of half-integers (for odd $q/\Delta_p$) or integers (for even $q/\Delta_p$),
\begin{align}
\begin{split}
\hat{e}_{j}\equiv &\hat{e}(q=j\Delta_p)
\\=\,&
    \frac{1}{2j \Delta_p} \sum_k \Big( 
    S((k+j)\Delta_p, k\Delta_p) \big(\frmn_{k}^\dagger \frmn^{}_{k+j} - \anfm^\dagger_{k}  \anfm^{}_{k+j} \big)
\\&
    -A((k+j)\Delta_p, k\Delta_p) \big(\frmn_{k}^\dagger \anfm^\dagger_{-k-j} + \frmn^{}_{-k} \anfm^{}_{k+j}\big) \Big)\,,
\end{split}
\end{align}
where summation of $k$ runs from $\max(-N_\Lambda, j-N_\Lambda)$ to $\min(N_\Lambda, N_\Lambda-j)$. The 
electric field energy is given by
\begin{align}
    \hat{H}_\mathrm{Sch}^\mathrm{int} =
 \frac{g_\mathrm{Sch}^2}{4\pi\,\Delta_p} \sum_{j=1}^{N_\Lambda} (\hat{e}_{j} \hat{e}_{-j} + \hat{e}_{-j} \hat{e}_{j}).
\end{align}
Similarly, we may obtain $\hat{\sigma}_{j}\equiv \hat{\sigma}(q=j\Delta_p)$ and 
\begin{align}
     \hat{H}_\mathrm{NJL}^\mathrm{int} =
g_\mathrm{NJL}^2 \frac{\Delta_p}{2\pi} \sum_{j=-N_\Lambda}^{N_\Lambda} \hat{\sigma}_{j} \hat{\sigma}_{-j}.
\end{align}

In both Schwinger and NJL models, momentum-lattice formalisms of the full Hamiltonian ($\hat{H}_\mathrm{NJL}$ or $\hat{H}_\mathrm{Sch}$), the total momentum ($\hat{P}$), and the total charge ($\hat{Q}$) operators commute with each other. Thus, one can diagonalize them simultaneously. Particularly, the momentum Fock basis are already eigenstates of $\hat{P}$ and $\hat{Q}$, and the $\hat{H}_\mathrm{NJL}$ and $\hat{H}_\mathrm{Sch}$ matrices are block-diagonalized by construction---each block corresponds to the same eigenvalue combination of $\hat{P}$ and $\hat{Q}$. 
Therefore, to find all the eignevalues and eigenvectors of the Hamiltonian matrices, one may diagonalized each block independently, which allows for a complete exact diagonalization (ED) for a relatively larger grid.

In this paper we perform ED for a system with $N_\Lambda = 4$, which corresponds to 9 momentum points, and equivalently, 18 lattice points. The dimension of the Hilbert space is $2^{18}=262144$, which is unrealistic for ED. Nevertheless, since we focus on the sub-sector with $Q=0$ and $P=0$ which has dimension $3368$. The full property can be studied with very high numerical efficiency. 

In the momentum-space lattice theory, the infrared(IR) and ultraviolet(UV) cut-offs are set by the discretization spacing ($\Lambda_\mathrm{IR} \equiv \Delta_p$) and the maximum momentum ($\Lambda_\mathrm{UV} \equiv N_\Lambda\,\Delta_p$), respectively. For better interpretation with the continuous theory, we set dimensional quantities (i.e., fermion mass and Schwinger coupling) to be between the IR and UV scales, $\Lambda_\mathrm{IR} < m, g_\mathrm{Sch} < \Lambda_\mathrm{UV}$. Thus, letting $g \equiv g_\mathrm{Sch}/\sqrt{2}$ be the energy unit, we set $\Delta_p = g/2$ and $m=0.6g$ and study the static feature of a thermal systems. The NJL coupling, which is unitless, is set to be $g_\mathrm{NJL}= 2$ so that the energy spectrum matches well with that of the Schwinger model.

\begin{figure}
    \centering
    \includegraphics[width=0.9\linewidth]{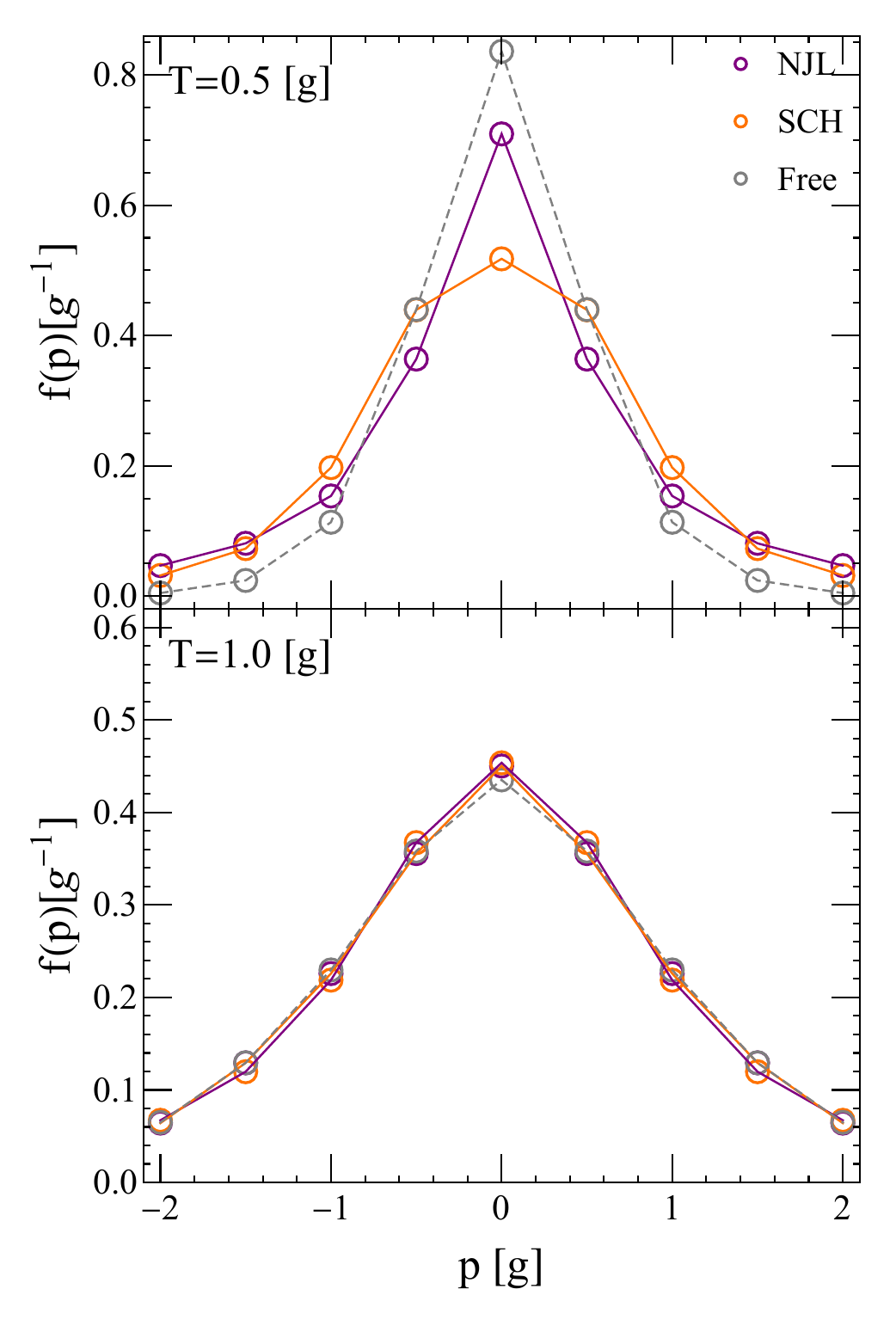}
    \caption{Equilibrium single-particle momentum distribution function for NJL (purple) and Schwinger (orange) models with comparison to interaction free distribution under temperatures $T=0.5g$ (upper) and $T=g$ (lower). \label{fig:thermal1}}
\end{figure}

\section{Single-particle momentum distribution}\label{sec:results1}
In this work, we are interested in the thermal equilibrium state, described by the density matrix $\hat{\rho}(T) = e^{-\hat{H}/T}/Z(T)$. $\hat{H}$ is the full Hamiltonian for Schwinger and NJL models, respectively, and $Z(T) \equiv \mathrm{tr}\big(e^{-\hat{H}/T}\big)$ ensures the normalization of density matrix. For later convenience, we denote the thermal expectation $\langle \hat{O} \rangle_T \equiv \mathrm{tr}\big(\hat{\rho}(T) \hat{O} \big)$. Note that the UV cut-off in the lattice theory sets an upper bound of the energy levels. For better comparison with continuous theories, one should focus on temperatures below the UV scale.\footnote{Finite temperature properties are save with respect to the IR cut-off---a lattice theory with proper parameter settings ($\Lambda_\mathrm{IR} < m, g_\mathrm{Sch} < \Lambda_\mathrm{UV}$) is expected to reproduce the ground-state and lower excitations. Therefore, equilibrium properties with low temperature ($T\lesssim \Lambda_\mathrm{IR}$) shall still be comparable with the continuous theory.}

\begin{figure}[!hbt]
    \centering
    \includegraphics[width=0.8\linewidth]{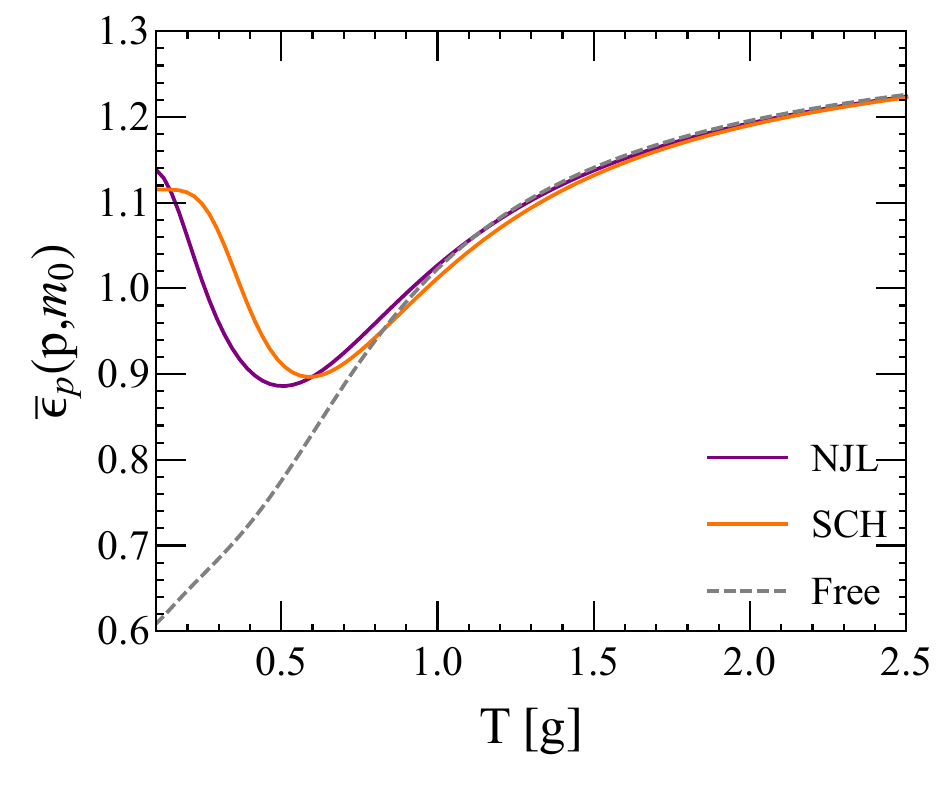}
    \caption{Kinetic energy per particle of interacting theories verse interaction free case. \label{fig:energy_per_particle}}
\end{figure}

\begin{figure}[!hbt]
    \centering
    \includegraphics[width=0.8\linewidth]{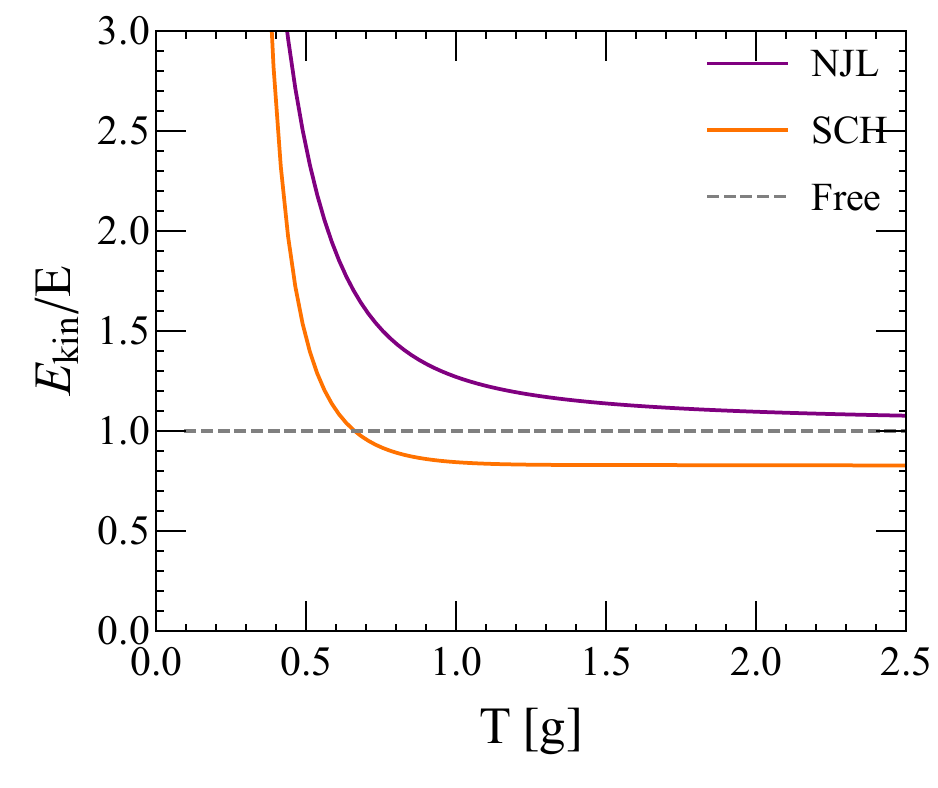}
    \caption{Free kinetic energy over the total energy as a function of temperature.\label{fig:kinetic_energy}}
\end{figure}
We first calculate the single-fermion momentum distribution $f(p;T) \equiv \langle \hat{N}(p) \rangle_T /\mathcal{N}(T)$, with $\hat{N}(p) \equiv \frmn^\dagger(p) \frmn^{}(p)$ being the fermion number operator and $\mathcal{N}(T)$ the normalization factor ensuring $\int f(p) \mathrm{d}p =1$. They are shown Fig.~\ref{fig:thermal1}, which compares the Schwinger and NJL results with the interaction free  distribution (i.e., taking $\hat{H} = \hat{H}_\mathrm{Dirac}$). Since the total charge and momentum of our discretized system are fixed to be $Q=0$ and $P=0$, the free distribution is deviated from Fermi -- Dirac distribution which is set to be
\begin{align}
\begin{split}
    &f_\mathrm{free}(p_n;T)
\\=&
    \frac{e^{-2\beta \epsilon_n} +e^{-\beta\epsilon_n}\sum_{ijk}e^{-\beta(\epsilon_{i}+\epsilon_{k}+\epsilon_{k})}\delta_{i+j+k+n,0}+\cdots}{1+\sum_{i}e^{-2\beta\epsilon_i}+\sum_{ijkl}e^{-\beta(\epsilon_{i}+\epsilon_{k}+\epsilon_{k}+\epsilon_{l})}\delta_{i+j+k+l,0}+\cdots}.
\end{split}
\end{align}
We take two temperature points, $T=0.5\,g$ and $1.0\,g$.
For the lower temperature, both NJL and Schwinger models exhibit momentum tails compared to the interaction free distribution. This owns to the fact that the attractive interaction in both models forms boundstates of fermion-antifermion pair--- i.e., the $\sigma$ meson in this single-flavor NJL model~\cite{Nambu:1961tp, Nishiyama:2001hx} and the well known scalar meson in Schwinger model~\cite{Schwinger:1962tp}---and the fermions carry extra kinetic energy associated with the relative motion within the boundstates. The in-boundstate motion of fermions makes the momentum distribution wider than that of the random thermal motion.
Meanwhile, one may observe distinguishable features between the NJL and Schwinger models, reflecting their difference in the inner structure of such boundstates. 

At higher temperature ($T=g$), we observe that the NJL and Schwinger models exhibit similar $f(p)$ i.e., the large momentum tail get suppressed. Single-particle distributions in the interacting theories are similar to a interaction-free distribution with a little bit lower temperature, which means that interactions between the spinors become subdominant compared to the kinetic collisions at high temperature, so that distinctions between the different interacting models become small. In this high-temperature perturbative limit, the fermions become quasi-free, and this could also be seen in the temperature-dependent kinetic energy per particle, defined as $\bar \epsilon_p = \int \mathrm{d}p\, \sqrt{m^2+p^2} f(p)$ while $m$ is the bare fermion mass. Results in Fig.~\ref{fig:energy_per_particle} show a higher kinetic energy of single fermion state than free ones below a certain temperature ($T \lesssim g$) and approaches to free particle in high temperature limit.

To further digest the effective degree of freedom in the medium, we compute the ratio between kinetic energy $E_\mathrm{kin} \equiv \langle \int \hat{H}_\mathrm{Dirac} \rangle_T$ and total energy $E \equiv \langle \hat{H}\rangle_T$. See Fig.~\ref{fig:kinetic_energy}. When the temperature is low, the interactive interaction ($E_\mathrm{int} \equiv \langle \hat{H}_\mathrm{int} \rangle_T = \langle \hat{H}\rangle_T - \langle \hat{H}_\mathrm{Dirac} \rangle_T$) has made the total energy significantly lower than the free kinetic energy. As $T$ increases above the interaction strength, contribution of interactions decreases rapidly
\footnote{We note that in the Schwinger model, interaction energy is carried by the electric field energy. It contributes positively at high temperature,  although remaining attractive.}, so that $E \approx E_\mathrm{kin}$. For such hot medium, the fermions and antifermions are asymptotically free and become good quasiparticles of the system. However, comparing the result of kinetic energy per particle and ratio between kinetic energy and total energy we would like to find that when the temperature is around $T\sim g$, the quasi-particle kinetic properties are almost the same while the information of interaction is still stored in the thermal quantities.

We may further connect the momentum distribution with the effective self-energy [$\Sigma(\omega,p)$], defined as the interaction correction to the fermion's propergator in momentum space, 
\begin{align}
\begin{split}
G^R(\omega,k) = \frac{1}{\omega - E_p - \Sigma(\omega,p)+i\epsilon}\,.
\end{split}
\end{align}
Here, $E_p = \sqrt{m^2+p^2}$ is the kinetic energy. The finite-temperature momentum distribution then reads
\begin{align}
\begin{split}
f(p,T) \propto \int \frac{\mathrm{d}\omega}{1+e^{\omega\,T}} \frac{-2\,\mathrm{Im} \Sigma}{(\omega-E_p - \mathrm{Re}\Sigma)^2 + (\mathrm{Im}\Sigma)^2}\,.
\end{split}
\end{align}
In low temperature region, the effective degrees of freedom emerges, contributes to a real $\Sigma$---corresponding to the static properties like thermal mass, and gives a long tail of higher momentum. In high temperature region with large scattering effect, the imaginary part of the self energy reach a more universal form regardless of the interacting details. 

\section{Connected four-momentum functions}\label{sec:results2}
\begin{figure}
    \centering
    \includegraphics[width=0.8\linewidth]{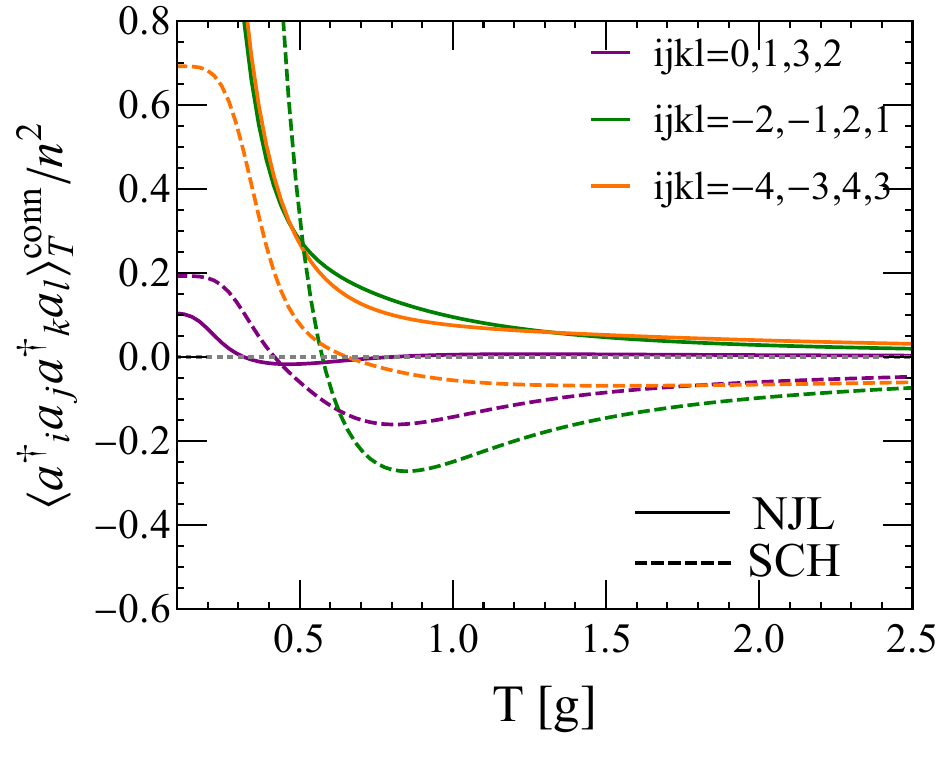}\\
    \includegraphics[width=0.8\linewidth]{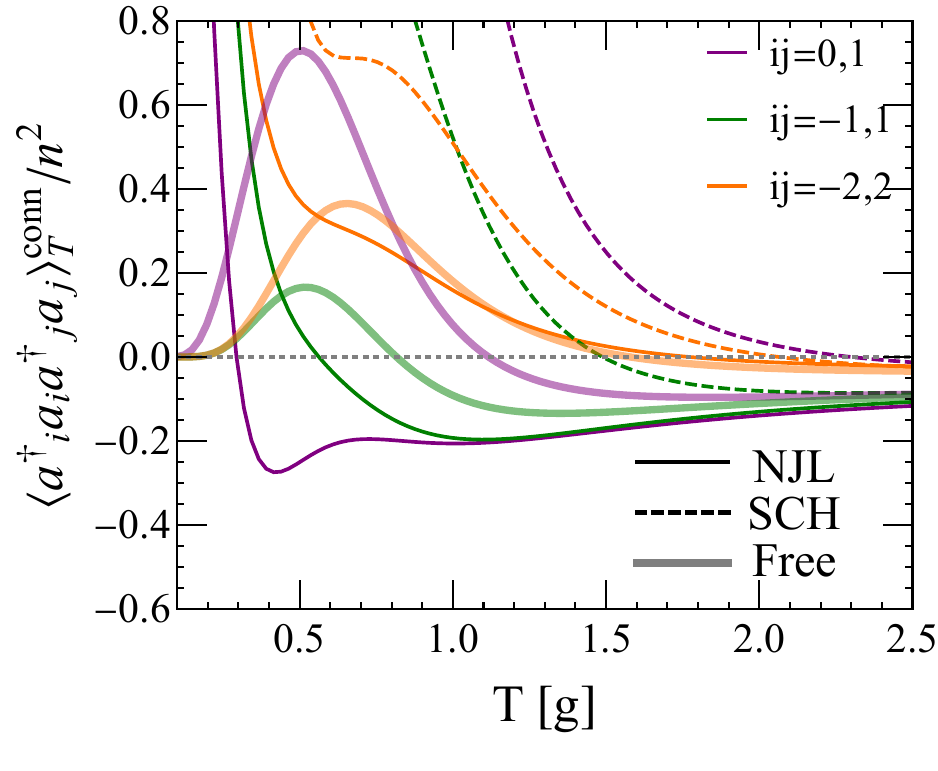}
    \caption{Thermal expectation of four-momentum correlator, $\langle \frmn^{\dagger}_{i}\frmn^{}_{j}\frmn^{\dagger}_{k}\frmn^{}_{l} \rangle_T$ (top panel) and connected two-point correlator $\langle \frmn^{\dagger}_{i}\frmn^{}_{i}\frmn^{\dagger}_{j}\frmn^{}_{j} \rangle_T-\langle \frmn^{\dagger}_{i}\frmn^{}_{i} \rangle_T\langle \frmn^{\dagger}_{j}\frmn^{}_{j}\rangle_T$ (bottom panel), scaled by averaged single-particle distribution squared, $\frac{1}{N}\sum_n \langle \frmn^{\dagger}_{n}\frmn^{}_{n} \rangle_T^2$. We compare the thermal expectations for $\hat{H}_\mathrm{NJL}$ (thin solid), $\hat{H}_\mathrm{Sch}$ (dashed), and $\hat{H}_\mathrm{Dirac}$ (thick opaque) theories. \label{fig:4pt2}}
\end{figure}

We then move on to calculate the connected four-momentum functions, which correspond to two-particle correlations. Denoting discrete fermion annihilation operators $\frmn^{}_{k} \equiv \sqrt{\frac{\Delta_p}{2\pi}}\,\frmn^{}(p=k\Delta_p)$, we defined the connected four-momentum functions as\footnote{Note that the other term in the disconnected contribution vanishes, are required by the charge conservation, $\langle \frmn^{\dagger}_{i} \frmn^{\dagger}_{k}\rangle_T = \langle \frmn^{}_{j}\frmn^{}_{l} \rangle_T = 0$. We also note that momentum conservation requires that $\langle \frmn^{\dagger}_{i}\frmn^{}_{j} \rangle_T = 0$ if $i\neq j$.}
\begin{align}
\begin{split}
    &\langle \frmn^{\dagger}_{i} \frmn^{}_{j} \frmn^{\dagger}_{k} \frmn^{}_{l} \rangle_{T}^{\mathrm{conn}} 
\equiv
    \langle \frmn^{\dagger}_{i} \frmn^{}_{j} \frmn^{\dagger}_{k} \frmn^{}_{l} \rangle_{T} 
\\&
    - 
    \langle \frmn^{\dagger}_{i} \frmn^{}_{j} \rangle_{T} \langle \frmn^{\dagger}_{k} \frmn^{}_{l} \rangle_{T}
    +\langle \frmn^{\dagger}_{i} \frmn^{}_{l} \rangle_{T} \langle \frmn^{\dagger}_{k} \frmn^{}_{j} \rangle_{T}.
\end{split}
\end{align}
The operator creates two fermions with momentum $i\,\Delta_p$ and $k\,\Delta_p$ and then annihilating two fermions with momentum $j\Delta_p$ and $l\Delta_p$. We have required that $i+k = j+l$ to fulfill total momentum conservation. We are interested in two types of combinations of $i,j,k,l$: first, $i=j \neq k=l$ so that the disconnected expectation corresponds to the two-particle correlation in momentum space $g(p,k) \equiv f(p,k) - f(p)f(k)$; and second, they are different from each other and the disconnected terms always vanish. These two types of four-momentum functions are shown in Fig.~\ref{fig:4pt2}, scaled by the momentum-averaged of single-particle distribution squared, $n^2 \equiv (\sum_k \langle \frmn^\dagger_k \frmn^{}_k \rangle_{T}^2) / (\sum_k 1)$ for a dimensionless comparison. The first observable detects the fractorization of particle density momentum distribution. 

In the interactionless case, with global constraints $Q=0$ and $P=0$, although there is no off-diagonal term in momentum Fock space, the connected two momentum correlator is non-zero, corresponding to a correlation of a classical probability distribution. The second observable further detect the quantum coherence sensitive to the off-diagonal term between different momentum configurations. And in this case, we could not find any correlation in interaction-free case. When we include the interaction, the things become more interesting. When we consider the connected two-momentum correlation function as shown in the lower panel of Fig.~\ref{fig:4pt2}, the correlators of the three types of interactions have the similar trend that in low temperature, the covariance of $n_p$ and $n_q$ is large and reduce with temperature. This correlation contains the mutual information of classical part and quantum part. However, for different momentum choice, we could also find that when we choose two momenta with opposite values, the correlation at high temperature are almost the same as interaction free case while for random two momenta, there are non-negligible difference between NJL and Schwinger case. We further come to the results of quantum coherence of momentum space in the upper panel of Fig.~\ref{fig:4pt2}, the correlation in Schwinger model keeps a considerable large value with $T\sim g$ in which temperature region, the kinetic energy of fermion is almost the same as the interaction free case. And the result of NJL model depends on the momenta choice. When we choose the random momenta that satisfy the total momentum change equals zero, but with out the parity in momentum space, the correlation would be much smaller than the parity case. 

Results are well fitted with the interaction versus thermal kinetic energy comparison. Correlations and connected four-momentum functions are comparable with the disconnected production of distributions when temperature is low and interaction dominates. As temperature increases, the connected-to-disconnected ratios decreases and tends to vanish when temperature becomes much higher than the interaction strength. 
We note that the four-momentum functions also measure the entanglement in momentum basis. 
Thus, small values in correlations and connected four-point functions reveal that the entangling sector of $\hat\rho_{jk}$ becomes negligible when interaction is much smaller than the thermal kinetic motion. The two-momentum quantum correlation functions $g(p,-p)$---defined as the difference between interacting theories and the interaction-free $\hat{H}_\mathrm{Dirac}$---decreases with $p$, indicating that weaker interaction with larger momentum exchange and causes smaller values in correlations/entanglements.

We end by noting the asymptotic free particle picture of both the NJL and Schwinger models at $T \approx g$, at which we observe the kinetic energies per particle of the interacting models match with that of the free Dirac spinors (Fig.~\ref{fig:energy_per_particle}), and particularly the similarity of the full single-particle distribution function (see, e.g., Fig.~\ref{fig:thermal1}, lower).  All of these seem to indicate that the system consist of free, disentangled fermions and antifermions. Nevertheless, the four-momentum function remain nonvanishing at such temperature (Fig.~\ref{fig:4pt2}), showing that the momentum correlation function remembers more microscopic details within a thermal system. In heavy-ion collisions, we anticipate that these correlations could be observed in connected energy-energy correlation between particles in a jet where quantum coherence has been preserved.

\section{Summary and Discussion}\label{sec:summary}
In this work, we first construct the matrix (quantum gate) representation of lattice spinor theories in the momentum space, which exactly conserve the total momentum, exhibit clear interpretation of momentum-related operators, and avoid doubling issue for coordinate-lattice spinor theories. Our construction of momentum-lattice spinor theories are ready to be simulated on both quantum and classical devices. Particularly, with the total momentum operator---together with the fermion charge operator---being conserved, the Hamiltonian is block-diagonalize in the momentum basis, and one may due with larger grids in the exact diagonalization treatment of such problems on classical computers.

With this framework, we study the finite-temperature properties of Schwinger and Nambu--Jona-Lasinio models, especially the single-particle and two-particle distribution functions.
For both models, we observe that the low-temperature properties are dominated by fermion-antifermion boundstates, which cause the high-momentum tail in single-particle distribution and large correlations in two-particle distribution and other connected four-momentum functions.
At high temperatures that is comparable with interaction, our results are consistent with quasi-free fermion/antifermion gas where the kinetic energy of the quasi-particles are almost the same as free Fermi particles. Thus, we anticipate that a kinetic theory of the fundamental particle should have been applicable when the thermal kinetic energy is comparable with the interaction. Nevertheless, the companion question---whether or not a kinetic theory that treats the boundstates as the emergent quasi-particles applies at low temperature---needs to be addressed in future.

The framework can be extended to studying theories. 
Notably, the high-momentum tail in single-particle momentum due to relative motion in boundstates are insensitive to the exact interaction, which sheds light on studying short-range correlations between neutron and proton in nuclei~\cite{Li:2022fhh, Si:2025eou}. This calls for studies with more realistic model that takes into account the isospin symmetry in the spinor field.
As a second example, one may take the ``small momentum assumptions'' and construct local theories in momentum space, so that the Tensor Network method could be applicable for larger grids.  

\section*{Acknowledgment}
This work is supported by NSFC under grant No. 12575143, by National Key Research and Development Program of China under Contract No. 2024YFA1610700, by Tsinghua University under grant Nos. 04200500123, 531205006, 533305009, by the Yantai University under grant NO.2226001, and by 2025/2026 INFN Research Grant Program grant No. 27076/2024. 

\bibliography{ref}

\end{document}